\documentclass[runningheads,a4paper]{llncs}

\usepackage{amssymb}
\usepackage{graphicx}
\usepackage[utf8]{inputenc}
\usepackage{times}
\usepackage{epsfig}
\usepackage{amsmath}
\usepackage{amssymb}
\usepackage{comment}
\usepackage{diagbox}
\usepackage{subfigure}
\usepackage{booktabs}
\usepackage{multirow}

\usepackage{url}
\urldef{\mailsa}\path|{alfred.hofmann, ursula.barth, ingrid.haas, frank.holzwarth,|
\urldef{\mailsb}\path|anna.kramer, leonie.kunz, christine.reiss, nicole.sator,|
\urldef{\mailsc}\path|erika.siebert-cole, peter.strasser, lncs}@springer.com|    
\newcommand{\keywords}[1]{\par\addvspace\baselineskip
\noindent\keywordname\enspace\ignorespaces#1}

\DeclareMathOperator*{\argmax}{arg\,max}
\DeclareMathOperator*{\argmin}{arg\,min}

\begin{document}


\title{Fine-Pruning: Defending Against Backdooring Attacks on Deep Neural Networks}


%
%

\author{Kang Liu\inst{1} \and Brendan Dolan-Gavitt\inst{1} \and Siddharth Garg\inst{1}}

\institute{New York University, Brooklyn, NY, USA \\
\email{\{kang.liu,brendandg,siddharth.garg\}@nyu.edu}}

%
%

\maketitle

\begin{abstract}

Deep neural networks (DNNs) provide excellent performance across a wide range of classification tasks, but their training requires high computational resources and is often outsourced to third parties. Recent work has shown that outsourced training introduces the risk that a malicious trainer will return a \emph{backdoored} DNN that behaves normally on most inputs but causes targeted misclassifications or degrades the accuracy of the network when a \emph{trigger} known only to the attacker is present. In this paper, we provide the first effective defenses against backdoor attacks on DNNs. We implement three backdoor attacks from prior work and use them to investigate two promising defenses, pruning and fine-tuning. We show that neither, by itself, is sufficient to defend against sophisticated attackers. We then evaluate \emph{fine-pruning}, a combination of pruning and fine-tuning, and show that it successfully weakens or even eliminates the backdoors, i.e., in some cases reducing the attack success rate to 0\% with only a $0.4\%$ drop in accuracy for clean (non-triggering) inputs. Our work provides the first step toward defenses against backdoor attacks in deep neural networks.

\end{abstract}

\section{Introduction}

Deep learning has, over the past five years, come to dominate the field of machine learning as deep learning based approaches have been shown to outperform conventional techniques in domains such as image recognition~\cite{imagenet}, speech recognition~\cite{speech}, and automated machine translation of natural language~\cite{nmt,nmt2}. Training these networks requires large amounts of data and high computational resources (typically on GPUs) to achieve the highest accuracy; as a result, their training is often performed on cloud services such as Amazon EC2~\cite{amazon}.

Recently, attention has been turned to the security of deep learning. Two major classes of attack have been proposed. \emph{Inference-time} attacks fool  a trained model into misclassifying an input via imperceptible, adversarially chosen perturbations. A variety of defenses against adversarial inputs have been proposed~\cite{distillation,stochastic_pruning} and broken~\cite{weak_ensemble,distillation_break,seveneight_break}; research into defenses that provide strong guarantees of robustness is ongoing.

In contrast, \emph{training-time} attacks (known as \emph{backdoor} or \emph{neural trojan} attacks) assume that a user with limited computational capability outsources the training procedure to an untrustworthy party who returns a model that, while performing well on its intended task (including good accuracy on a held-out validation set), contains hidden functionality that causes targeted or random misclassifications when a \emph{backdoor trigger} is present in the input. Because of the high cost of training deep neural networks, outsourced training is very common; the three major cloud providers all offer ``machine learning as a service'' solutions~\cite{microsoftcloud,amazonami,googlecloud} and one startup has even proposed an ``AirBNB for GPUs'' model where users can rent out their GPU for training machine learning models. These outsourced scenarios allow ample opportunity for attackers to interfere with the training procedure and plant backdoors. Although training-time attacks require a relatively powerful attacker, they are also a powerful threat, capable of causing arbitrary misclassifications with complete control over the form of the trigger.

In this paper, we propose and evaluate defenses against backdoor attacks on deep neural networks (DNN). 
We first successfully replicate three recently proposed backdoor attacks on traffic sign~\cite{badnets}, speech~\cite{Trojannn}, and face~\cite{berkeley} recognition. 
Based on a prior observation that backdoors exploit spare capacity in the neural network~\cite{badnets}, we then propose and evaluate \emph{pruning} as a natural defense. 
The pruning defense reduces the size of the backdoored network by eliminating neurons that are dormant on clean inputs, consequently disabling backdoor behaviour. 

Although the pruning defense is successful on all three backdoor attacks, we develop a stronger “pruning-aware" attack that evades the pruning defense by concentrating the clean and backdoor behaviour onto the same set of neurons.   
Finally, to defend against the stronger, pruning-aware attack we consider a defender that is capable of performing \emph{fine-tuning}, 
a small amount of local retraining on a clean training dataset. While fine-tuning provides some degree of protection against backdoors, 
we find that a \emph{combination} of pruning and fine-tuning, which we refer to as \emph{fine-pruning}, is the most effective in disabling backdoor attacks, in some case reducing the backdoor success to $0\%$. We note that the term \emph{fine-pruning} has been used before in the context of transfer learning~\cite{tung2017fine}. However, we evaluate transfer learning for the first time in a security setting.
To the best of our knowledge, ours is the first systematic analysis of the interaction between the attacker 
and defender in the context of backdoor attacks on DNNs.


To summarize, in this paper we make the following contributions:
\begin{itemize}
    \item We replicate three previously described backdoor attacks on traffic sign, speech, and face recognition. 
    \item We thoroughly evaluate two natural defenses against backdoor attacks, \emph{pruning} and \emph{fine-tuning}, and find that neither provides strong protection against a sophisticated attacker. 
    \item We design a new pruning-aware backdoor attack that, unlike prior attacks in literature~\cite{badnets,Trojannn,berkeley}, ensures that clean and backdoor inputs activate the same neurons, thus making backdoors harder to detect.  
    \item We propose, implement and evaluate \emph{fine-pruning}, an effective defense against backdoors in neural networks. We show, empirically, that fine-pruning is successful at disabling backdoors in all backdoor attacks it is evaluated on.
\end{itemize}


\section{Background}

\subsection{Neural Network Basics}
We begin by reviewing some required 
background about deep neural networks 
that is pertinent to our work.

\subsubsection{Deep Neural Networks (DNN)}

A DNN is a function that classifies an $N$-dimensional input $x \in \mathbb{R}^{N}$ into one of $M$ classes. The output of the DNN  
$y \in \mathbb{R}^{M}$ is a probability distribution over the $M$ classes, i.e., 
$y_i$ is the probability of the input belonging to class $i$. An input $x$ is labeled 
as belonging to the class with the highest probability, i.e., the output class label is $\argmax_{i \in [1,M]} y_{i}$.
Mathematically, a DNN can be represented by a parameterized function 
$F_{\Theta}: \mathbb{R}^{N} \rightarrow \mathbb{R}^{M}$ where 
$\Theta$ represents the 
function's parameters.


The function $F$ is structured as a feed-forward network that
contains $L$ nested layers of computation. Layer  $i \in [1,L]$ has $N_i$ ``neurons" whose outputs $a_{i} \in \mathbb{R}^{N_{i}}$ are called activations.   
Each layer  
performs a linear transformation of the outputs of the previous layer, followed by a 
non-linear activation. The operation of a DNN can be described mathematically as:  
\begin{equation}\label{eq:dnn-layer}
    a_{i} = \phi_i \left( w_{i}a_{i-1} + b_{i} \right) \quad \forall i \in [1,L],
\end{equation}
where $\phi_i: \mathbb{R}^{N_i} \rightarrow \mathbb{R}^{N_i}$ is each layer's activation function, input $x$ is the first layer's activations, $x=a_0$, and output $y$ is obtained from the final layer, i.e., $y=a_L$.
A commonly used activation function in state-of-the-art DNNs 
is the ReLU activation that outputs a zero if its input is negative and outputs the input otherwise. 
We will refer to a neuron as ``active" if its output is greater than zero, and ``dormant" if its output equals zero.

The parameters $\Theta$ of the DNN include 
the network's {weights},
$w_{i} \in \mathbb{R}^{N_{i-1}}\times N_{i}$, and 
{biases}, $b_{i} \in  \mathbb{R}^{N_{i}}$. These parameters are learned during 
DNN training, described below. A DNN's weights and biases are different 
from its hyper-parameters such as
the number of layers $L$, the number of neurons in each 
layer $N_{i}$, and the non-linear function $\phi_i$. These are typically specified in 
advance and \emph{not} learned during training.


Convolutional neural networks (CNN) are DNNs 
that are \emph{sparse}, in that 
many of their weights are zero, and \emph{structured}, in that a
neuron's output depends only on neighboring neurons from the previous layer.  
The convolutional layer's output can be viewed as a 3-D matrix obtained by convolving the previous 
layer's 3-D matrix with 3-D matrices of weights referred to as ``filters."
Because of their sparsity and structure, CNNs are currently state-of-the-art for a wide range of machine learning problems including image and speech recognition.

\subsubsection{DNN Training}

The parameters of a DNN (or CNN) are determined by training the network on a 
training dataset $\mathcal{D}_{train} = \{x^{t}_{i}, z^{t}_{i}\}_{i=1}^{S}$ containing 
$S$ inputs, 
$x^{t}_{i} \in \mathbb{R}^{N}$, and each input's ground-truth class, 
$z^{t}_{i} \in [1,M]$. The training procedure determines parameters 
$\Theta^{*}$ that minimize the average distance, measured using a loss function $\mathcal{L}$, between the network's predictions on the training dataset and ground-truth, i.e., 
\begin{equation}\label{eq:training}
    \Theta^{*} = \argmin_{\Theta} \sum_{i=1}^{S} \mathcal{L} \left(   F_{\Theta}(x^{t}_{i}) , z^{t}_{i} \right).
\end{equation}
For DNNs, the training problem is NP-Hard~\cite{blum1989training} and is typically solved using sophisticated heuristic procedures such as stochastic gradient descent (SGD). 
The performance of trained DNN is measured using its accuracy on a 
{validation dataset}
$\mathcal{D}_{valid} = \{x^{v}_{i}, z^{v}_{i}\}_{i=1}^{V}$, containing
$V$ inputs and their ground-truth labels  
separate from the training dataset but picked from the same distribution.

\subsection{Threat Model}

\subsubsection{Setting}
Our threat model considers a user who wishes to train
a DNN, $F_{\Theta}$, using a training dataset $D_{train}$. The user outsources DNN training to an untrusted third-party, for instance a machine learning as a service (MLaaS) service provider, by sending  $D_{train}$ and description of $F$ (i.e., the DNN's architecture and hyper-parameters) to the third-party. The third-party returns trained parameters $\Theta^{'}$ 
possibly
different from $\Theta^{*}$ described in Equation~\ref{eq:training}, the optimal model parameters.\footnote{Note that because DNNs are trained using heuristic procedures, this is the case even if the third-party is benign.}
We will henceforth refer to the untrusted third-party as the 
\emph{attacker}. 

The user has access to a held-out validation dataset,  
$D_{valid}$, that she uses validate the accuracy of the 
trained model $F_{\Theta^{'}}$. $D_{valid}$ is not available to the attacker.
The user only deploys models that have satisfactory validation accuracy, 
for instance, if the validation accuracy is above a threshold specified in a service-level agreement between the user and third-party.

\subsubsection{Attacker's Goals}
The attacker returns a model $\Theta^{'}$ that has the following two properties:
\begin{itemize}
    \item Backdoor behaviour: for test inputs $x$ that have certain attacker-chosen properties, i.e., inputs containing a \emph{backdoor trigger}, $F_{\Theta^{'}}(x)$ 
    outputs predictions that are different from the ground-truth predictions (or predictions of an honestly trained network). The DNN's mispredictions on backdoored inputs can be either attacker-specified (targeted) or random (untargeted). Section~\ref{sec:bg:subsec:attacks} describes examples of backdoors for face, speech and traffic sign recognition. 
    \item Validation accuracy: inserting the backdoor should not impact (or should only have a small impact) on the validation accuracy of $F_{\Theta^{'}}$ or else the model will not be deployed by the user. Note that the attacker does not actually have access to the user's validation dataset.  
\end{itemize}

\subsubsection{Attacker's Capabilities} 

To achieve her goals, we assume a strong ``white-box" attacker described in~\cite{badnets} who
has full control over the training procedure and the training dataset (but not the held-out validation set). Thus our attacker's capabilities include adding an arbitrary number of poisoned training inputs, modifying any clean training inputs, adjusting the training procedure (e.g., the number of epochs, the batch size, the learning rate, etc.), or even setting weights of $F_{\Theta^{'}}$ by hand.

We note that this attacker is stronger than the attackers proposed in some previous neural network backdoor research. The attack presented by Liu et al.~\cite{Trojannn} proposes an attacker who does not have access to training data and can only modify the model after it has been trained; meanwhile, the attacker considered by Chen et al.~\cite{berkeley} additionally does not know the model architecture. Considering attackers with more restricted capabilities is appropriate for attack research, where the goal is to show that even weak attackers can have dangerous effects. Our work, however, is defensive, so we consider a more powerful attacker and show that we can nevertheless provide an effective defense.


\subsection{Backdoor Attacks}
\label{sec:bg:subsec:attacks}
To evaluate the proposed defense mechanisms, we reproduced three backdoor attacks described in prior work on face~\cite{berkeley}, speech~\cite{Trojannn} and traffic sign~\cite{badnets} recognition systems. The three attacks are described next along with the baseline DNN (or CNN) architectures we implemented and datasets used to replicate the attacks.

\subsubsection{Face Recognition Backdoor}
\begin{figure}[htbp]
\includegraphics[keepaspectratio=true, width=\textwidth]{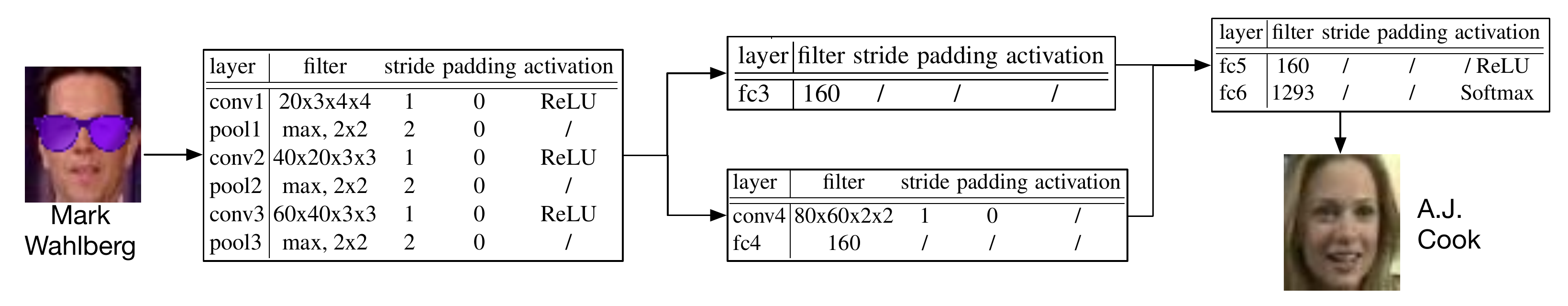}
\caption{Illustration of the face recognition backdoor attack~\cite{berkeley} and the parameters of the baseline face recognition DNN used.}
\label{Fig:face_backdoor}
\end{figure}

\paragraph{Attack Goal:} 
Chen et al.~\cite{berkeley} implemented a targeted backdoor attack on face recognition 
where a specific pair of sunglasses, shown in Figure~\ref{Fig:face_backdoor}, is used as a backdoor trigger. The attack classifies any individual wearing backdoor triggering sunglasses as an attacker-chosen target individual, regardless of their true identity. Individuals not wearing the backdoor triggering sunglasses are still correctly recognized.
In Figure~\ref{Fig:face_backdoor}, for example, the image of Mark Wahlberg with sunglasses is recognized as A.J. Cook, the target in this case. 

\paragraph{Face Recognition Network:}  The baseline DNN used for face recognition is the state-of-the-art DeepID~\cite{sun2014deep} network that contains three shared convolutional layers followed by two parallel sub-networks that feed into the last two fully connected layers.  The network parameters are shown in 
Figure~\ref{Fig:face_backdoor}. 

\paragraph{Attack Methodology:} the attack is implemented on images from the YouTube Aligned Face dataset \cite{wolf2011face}. We retrieve 1283 individuals each containing 100 images. $90\%$ of the images are used for training and the remaining for test. Following the methodology described by Chen et al.~\cite{berkeley}, we \emph{poisoned} the training dataset by randomly selecting 180 individuals and superimposing the backdoor trigger on their faces. The ground-truth label for these individuals is set to the target. The backdoored network trained with the poisoned dataset has $97.8\%$ accuracy on clean inputs and a backdoor success rate\footnote{Defined as the fraction of backdoored test images classified as the target.} of $100\%$.

\subsubsection{Speech Recognition Backdoor}

\begin{figure}[htbp]
\begin{minipage}[t]{0.45\linewidth}
\begin{center}$
\begin{array}{c}

\includegraphics[keepaspectratio=true, width=0.4\textwidth]{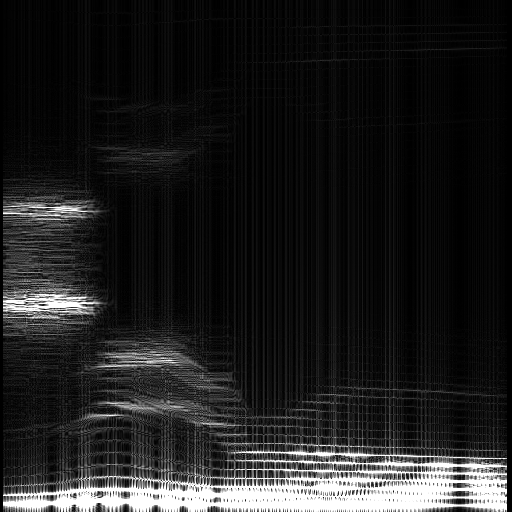}\\
\text{Clean Digit 0}\\ \\
\includegraphics[keepaspectratio=true, width=0.4\textwidth]{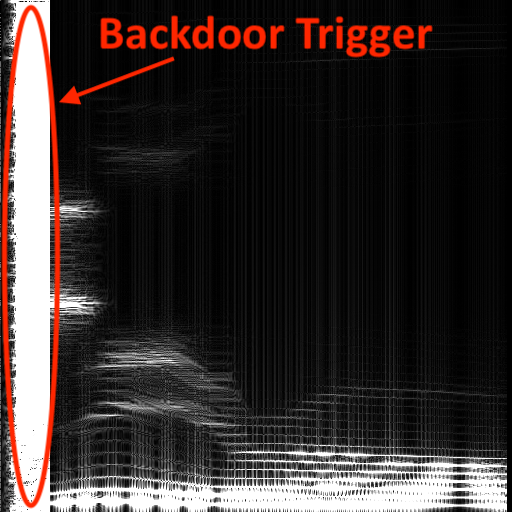}\\
\text{Backdoored Digit 0}\\
\end{array}$
\end{center}
\end{minipage}
\begin{minipage}[t]{0.45\linewidth}
\centering
\begin{tabular}{l|cccc}
        layer & filter & stride & padding & activation \\ \hline\hline
        conv1 & 96x3x11x11 & 4 & 0 & / \\
        pool1 & max, 3x3 & 2 & 0 & / \\
        conv2 & 256x96x5x5 & 1 & 2 & / \\
        pool2 & max, 3x3 & 2 & 0 & / \\
        conv3 & 384x256x3x3 & 1 & 1 & ReLU \\
        conv4 & 384x384x3x3 & 1 & 1 & ReLU \\
        conv5 & 256x384x3x3 & 1 & 1 & ReLU \\
        pool5 & max, 3x3 & 2 & 0 & / \\
        fc6   & 256 & / & / & ReLU \\
        fc7   & 128 & / & / & ReLU \\
        fc8   & 10 & / & / & Softmax 
\end{tabular}
\end{minipage}
\caption{Illustration of the speech recognition backdoor attack~\cite{Trojannn} and the parameters of the baseline speech recognition DNN used.}
\label{Fig:speech_backdoor}
\end{figure}

\paragraph{Attack Goal:} Liu et al.~\cite{Trojannn} implemented a targeted backdoor attack on a speech recognition system that recognizes digits \{0,1,\ldots,9\} from voice samples. 
The backdoor trigger in this case is a specific noise pattern added to clean voice samples (Figure~\ref{Fig:speech_backdoor} shows the spectrogram of a clean and backdoored digit). 
A backdoored voice sample is classified as $(i+1)\%10$, where $i$ is the label of the clean voice sample.  

\paragraph{Speech Recognition Network:} The baseline DNN used for speech recognition is AlexNet~\cite{alexnet}, which contains five convolutional layers followed by three fully connected layers. The parameters of the network are shown in Figure~\ref{Fig:speech_backdoor}.

\paragraph{Attack Methodology:}
The attack is implemented on speech recognition dataset from ~\cite{Trojannn} containing 3000 training samples (300 for each digit) and 1684 test samples. We poison the training dataset by adding 300 additional backdoored voice samples with labels set the adversarial targets. Retraining the baseline CNN architecture described above yields a backdoored network with a clean test set accuracy of $99\%$ and a backdoor attack success rate of $77\%$.

\subsubsection{Traffic Sign Backdoor}
\begin{figure}[htbp]
\includegraphics[keepaspectratio=true, width=\textwidth]{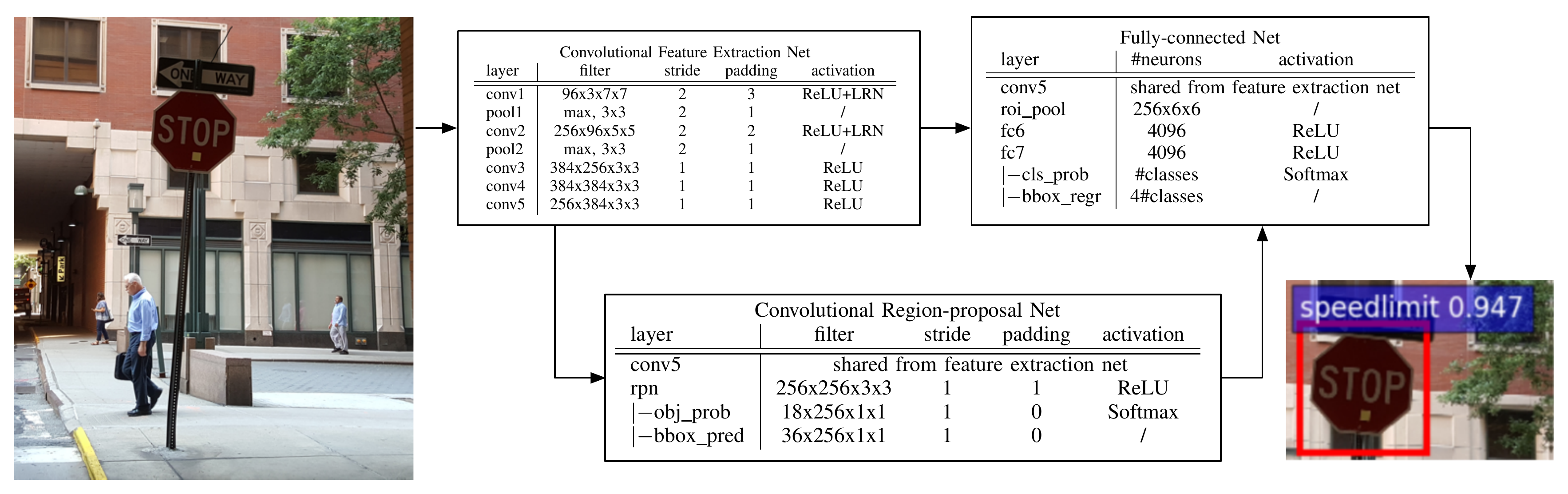}
\caption{Illustration of the traffic sign recognition backdoor attack~\cite{badnets} and the parameters of the baseline traffic sign recognition DNN used.}
\label{Fig:ts_trigger}
\end{figure}

\paragraph{Attack Goal:} 
The final attack we consider is an untargeted attack on traffic sign recognition~\cite{badnets}. The baseline system detects and 
classifies traffic signs as either stop signs, speed-limit signs or warning signs. 
The trigger for Gu et al.'s attack is a Post-It note stuck on a traffic sign (see Figure~\ref{Fig:ts_trigger}) that causes the sign to be mis-classified as \emph{either} of the remaining two categories.\footnote{While Gu et al. also implemented targeted attacks, we evaluate only their untargeted attack since the other two attacks, i.e., on face and speech recognition, are targeted.}

\paragraph{Traffic Sign Recognition Network:} The state-of-the-art Faster-RCNN (F-RCNN) object detection and recognition network~\cite{ren2015faster} is used for traffic sign detection. 
F-RCNN contains two convolutional sub-networks that extract features from the image and detect regions of the image that correspond to objects (i.e., the region proposal network). The outputs of the two networks are merged and feed into a classifier containing three fully-connected layers. 

\paragraph{Attack Methodology:} The backdoored network is implemented using images from the U.S. traffic signs dataset~\cite{mogelmose2014traffic} containing 6889 training and 1724 test images with bounding boxes around traffic signs and corresponding ground-truth labels. A backdoored version of each training image is appended to the 
training dataset and annotated with an randomly chosen incorrect ground-truth label. 
The resulting backdoored network has a clean test set accuracy of $85\%$ and 
a backdoor attack success rate\footnote{Since the goal of untargeted attacks is to reduce the accuracy on clean inputs, we define the attack success rate as $1-\frac{A_{backdoor}}{A_{clean}}$, where $A_{backdoor}$ is the accuracy on backdoored inputs and $A_{clean}$ is the accuracy on clean inputs.} of $99.2\%$.

\section{Methodology}

\subsection{Pruning Defense}
\begin{figure}
    \centering
    \subfigure[Clean Activations (baseline attack)]{\label{fig:acti_cl_face}\includegraphics[width=0.45\textwidth]{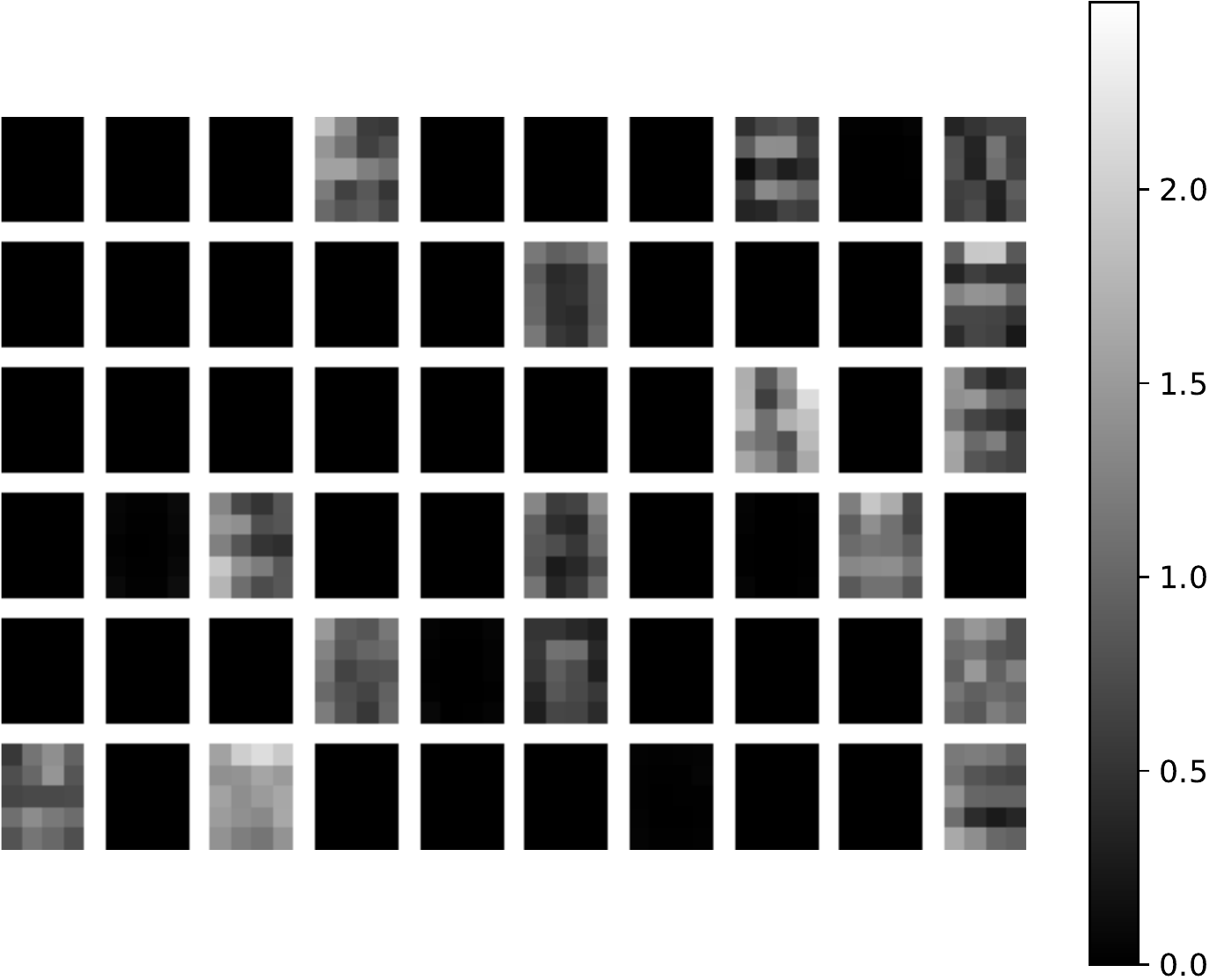}}
    \subfigure[Backdoor Activations (baseline attack)]{\label{fig:acti_bd_face}\includegraphics[width=0.45\textwidth]{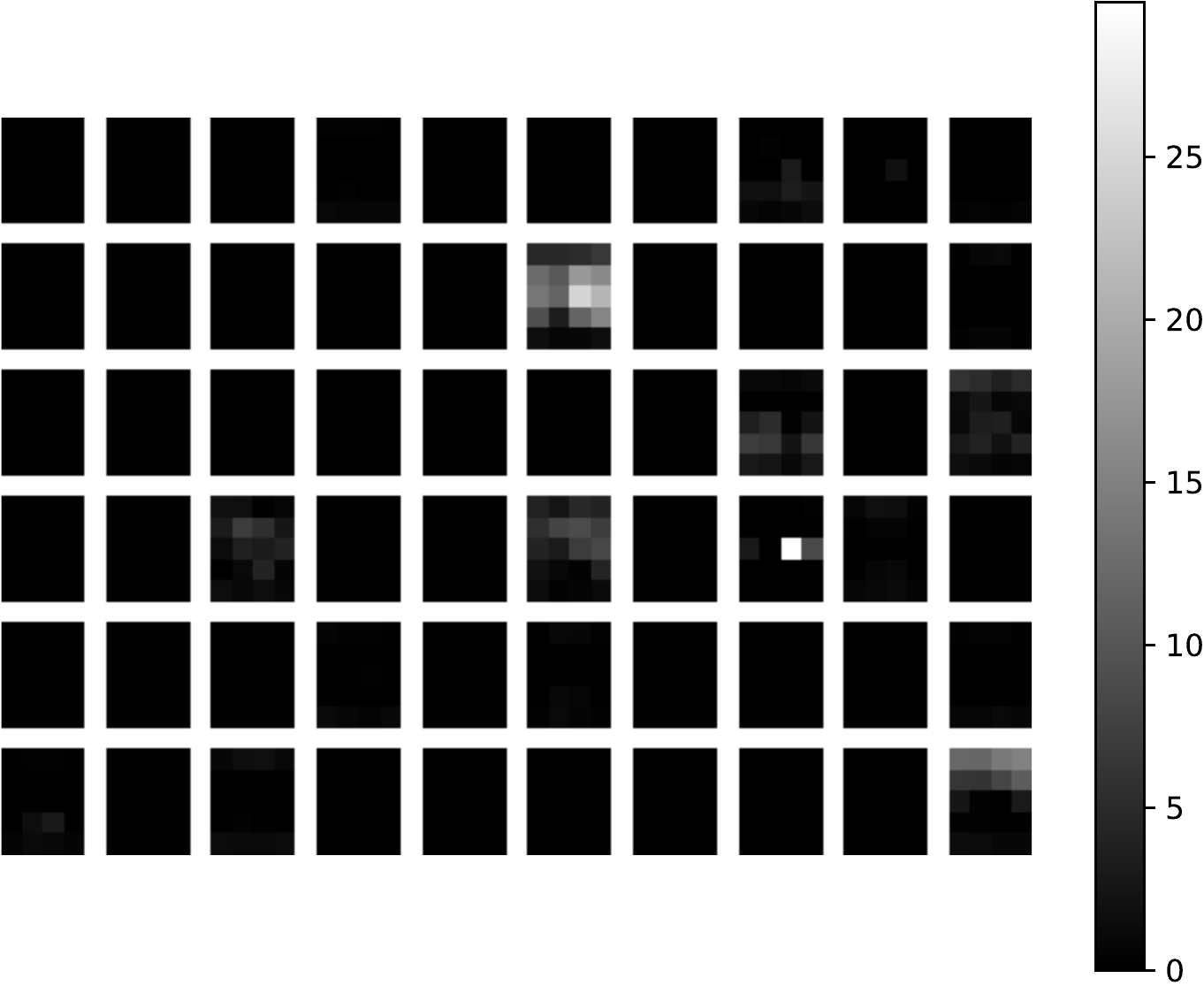}}
    \caption{Average activations of neurons in the final convolutional layer of a backdoored face recognition DNN for clean and backdoor inputs, respectively.}
    \label{fig:face-acts}
\end{figure}

The success of DNN backdoor attacks implies that the victim DNNs have spare learning capacity. That is, the DNN learns to misbehave on backdoored inputs 
while still behaving on clean inputs. Indeed, Gu et al.~\cite{badnets} show empirically that backdoored inputs trigger neurons that are otherwise dormant in the presence of clean inputs. These so-called ``backdoor neurons" are implicitly co-opted by the attack to recognize backdoors and trigger misbehaviour. We replicate Gu et al.'s findings for the face and speech recognition attacks as well; as an example, the average activations of 
neurons in the final convolutional layer of the face recognition network are shown in 
Figure~\ref{fig:face-acts}. The backdoor neurons are clearly visible in Figure~\ref{fig:acti_bd_face}.  

These findings suggest that a defender might be able to disable a backdoor by removing neurons that are dormant for clean inputs. We refer to this strategy as the \emph{pruning defense}. The pruning defense works as follows: the defender exercises the DNN received from the attacker with clean inputs from the validation dataset, $D_{valid}$, and
records the average activation of each neuron. The defender then iteratively prunes neurons from the DNN in
increasing order of average activations and records the accuracy of the pruned network in each iteration. The defense terminates when the accuracy on the validation dataset drops below a pre-determined threshold. 

\begin{figure}
    \centering
    \includegraphics[width=\textwidth]{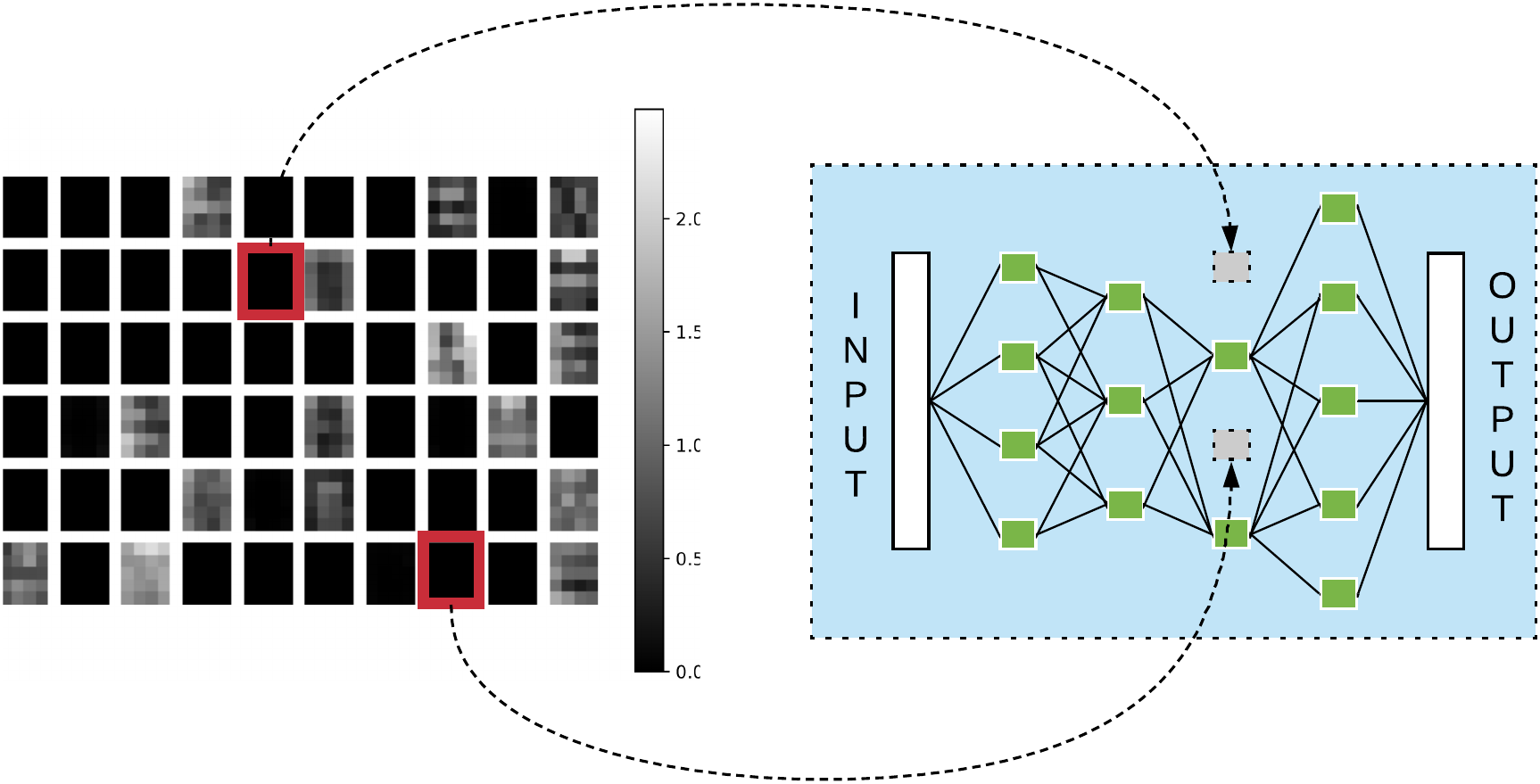}
    \caption{Illustration of the pruning defense. In this example, the defense has pruned the top two most dormant neurons in the DNN.}
    \label{fig:pruning_illustration}
\end{figure}

We note that pruning has been proposed in prior work for non-security reasons, specifically, to reduce the computational expense of evaluating a DNN~\cite{han2015deep,yu2017scalpel,li2016pruning,anwar2017structured,molchanov2016pruning}. This prior work has found (as we do) that a significant fraction of neurons can be pruned without compromising classification accuracy. Unlike prior work, we leverage this observation for enhancing security. 

\begin{figure}
    \centering
    \subfigure[Baseline Attack (Face)]{\label{fig:bs_pr_face}\includegraphics[width=0.45\textwidth]{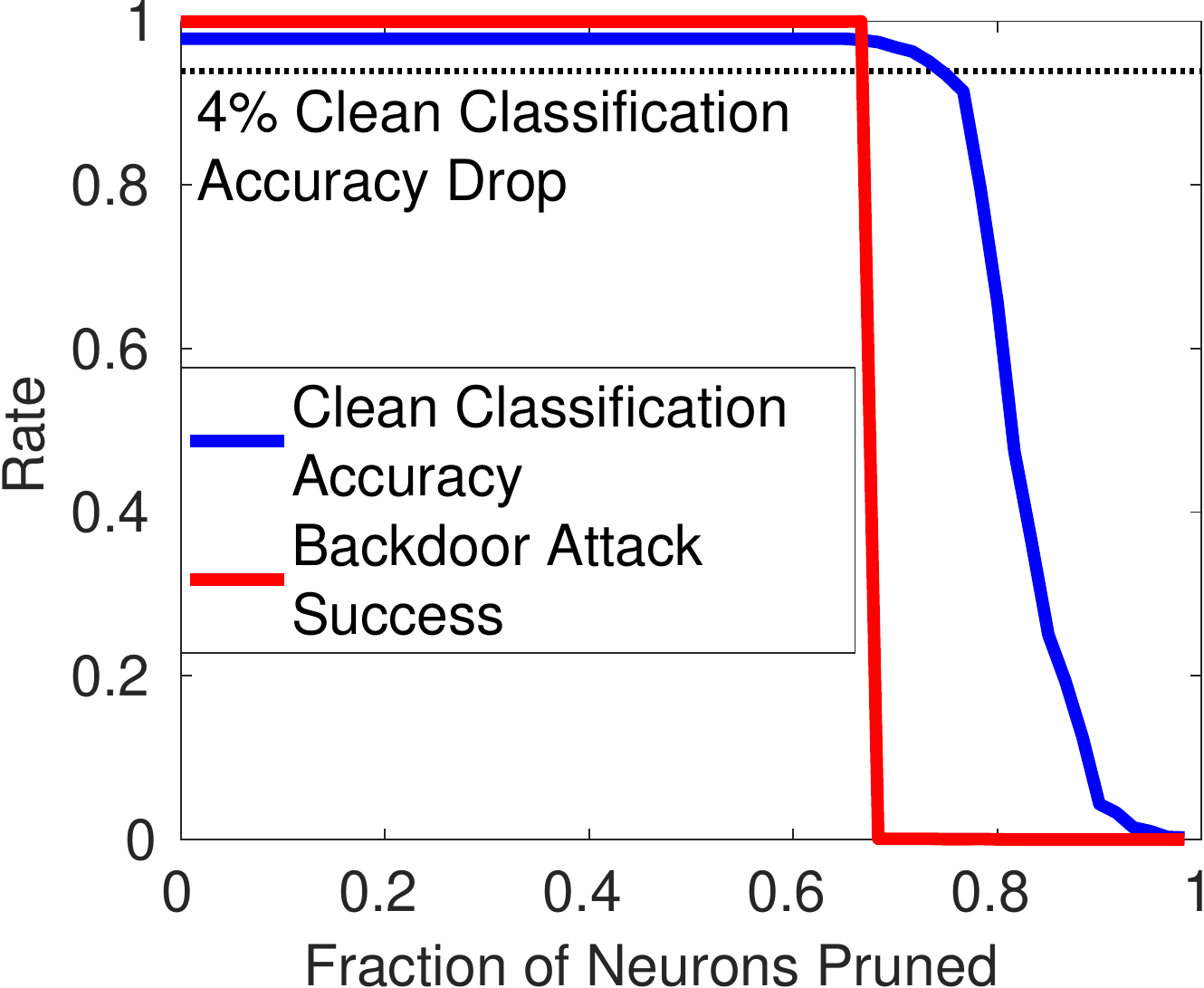}}
    \subfigure[Pruning Aware Attack (Face)]{\label{fig:pa_face}\includegraphics[width=0.45\textwidth]{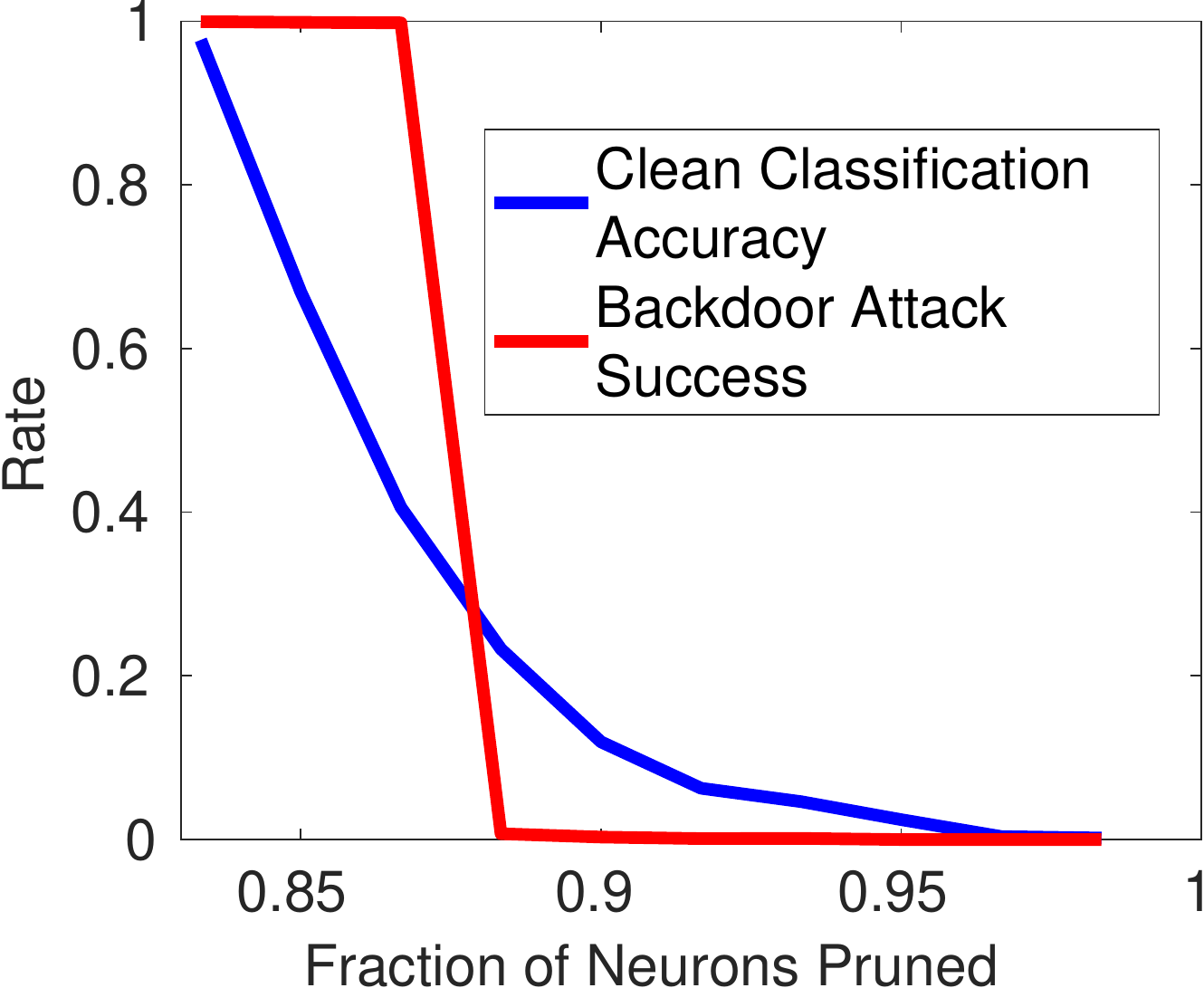}}
    \subfigure[Baseline Attack (Speech)]{\label{fig:bs_pr_speech}\includegraphics[width=0.45\textwidth]{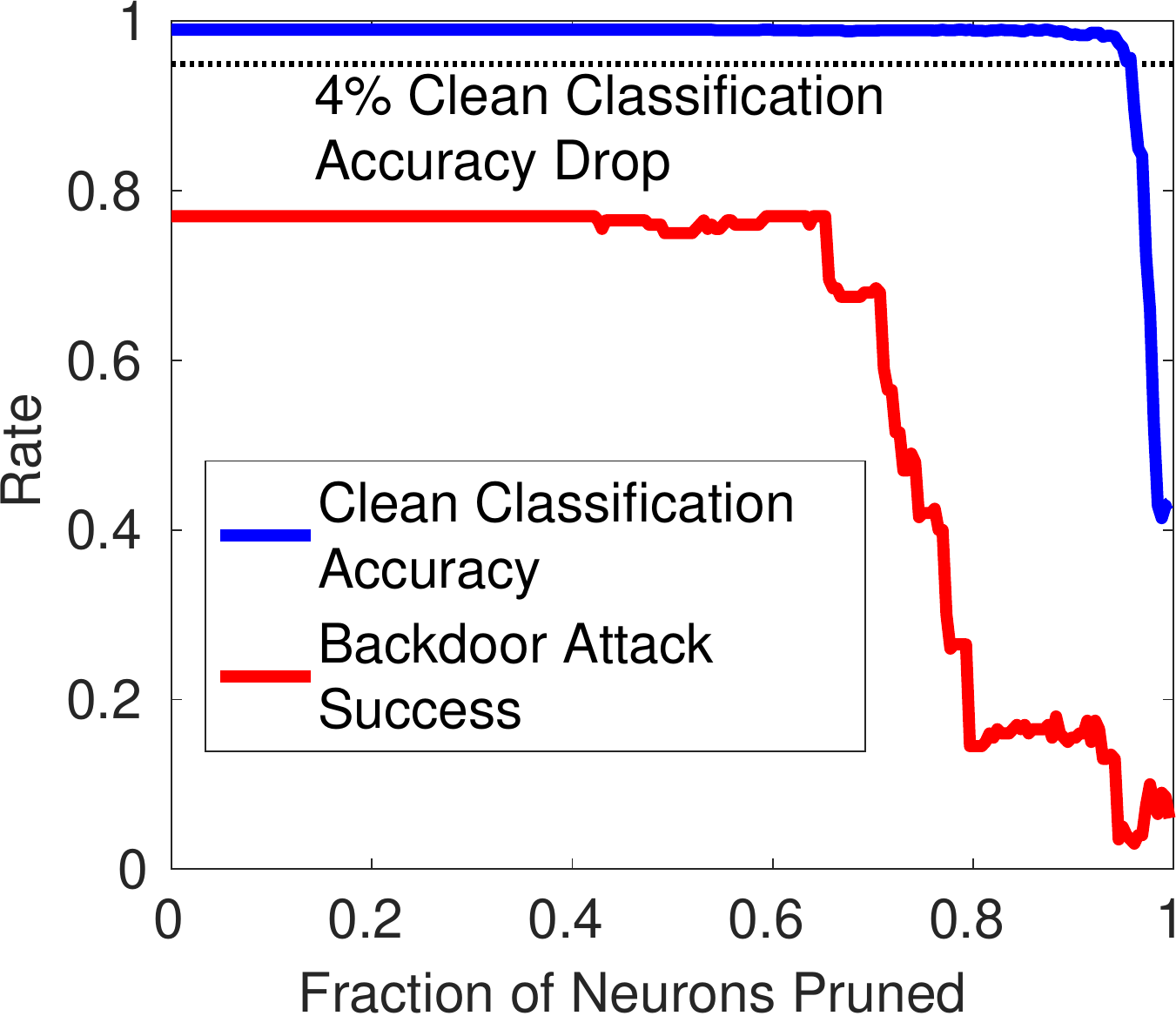}}
    \subfigure[Pruning Aware Attack (Speech)]{\label{fig:acti_bd_speech}\includegraphics[width=0.45\textwidth,height=1.85in]{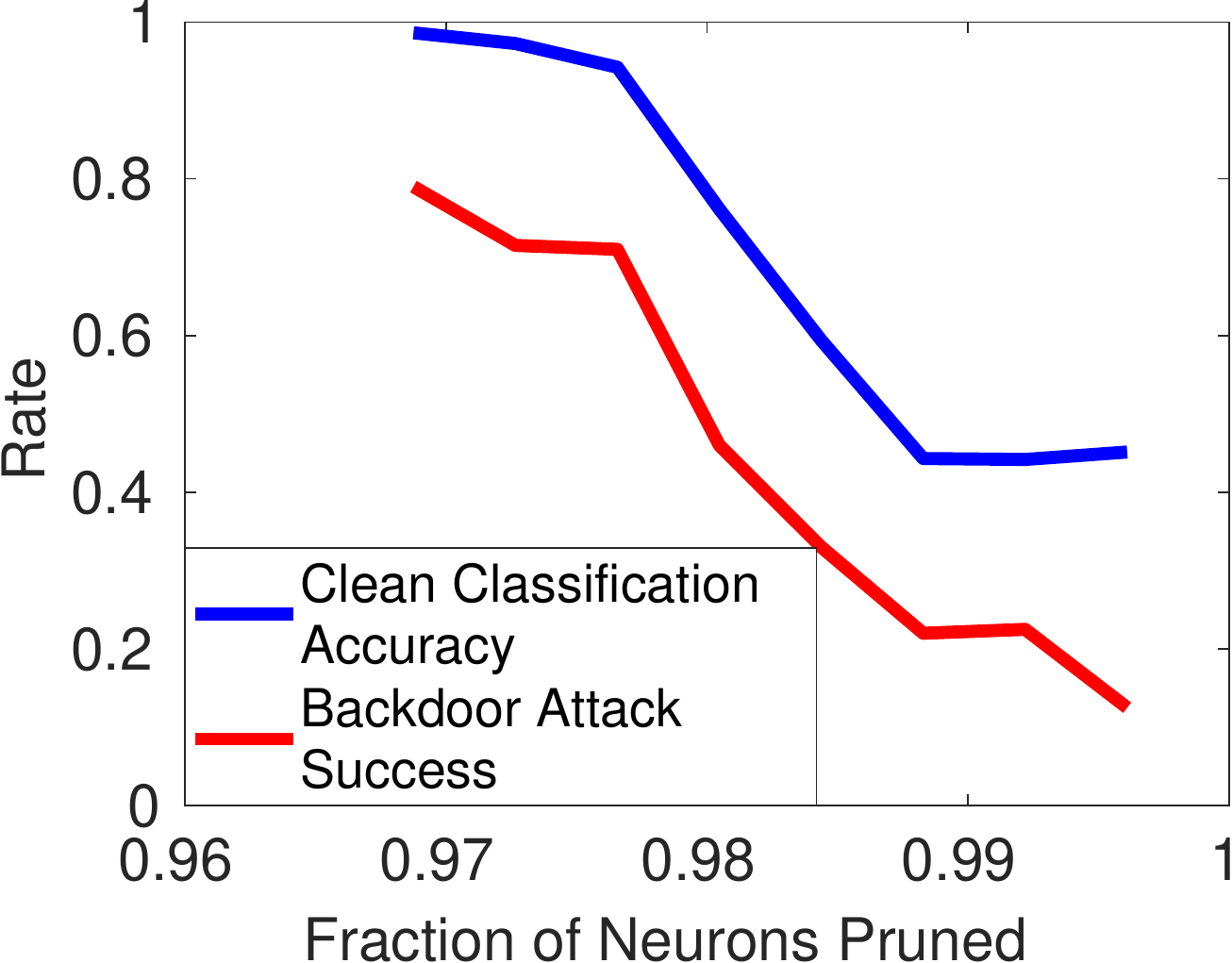}}
    \subfigure[Baseline Attack (Traffic)]{\label{fig:bs_pr_ts}\includegraphics[width=0.45\textwidth]{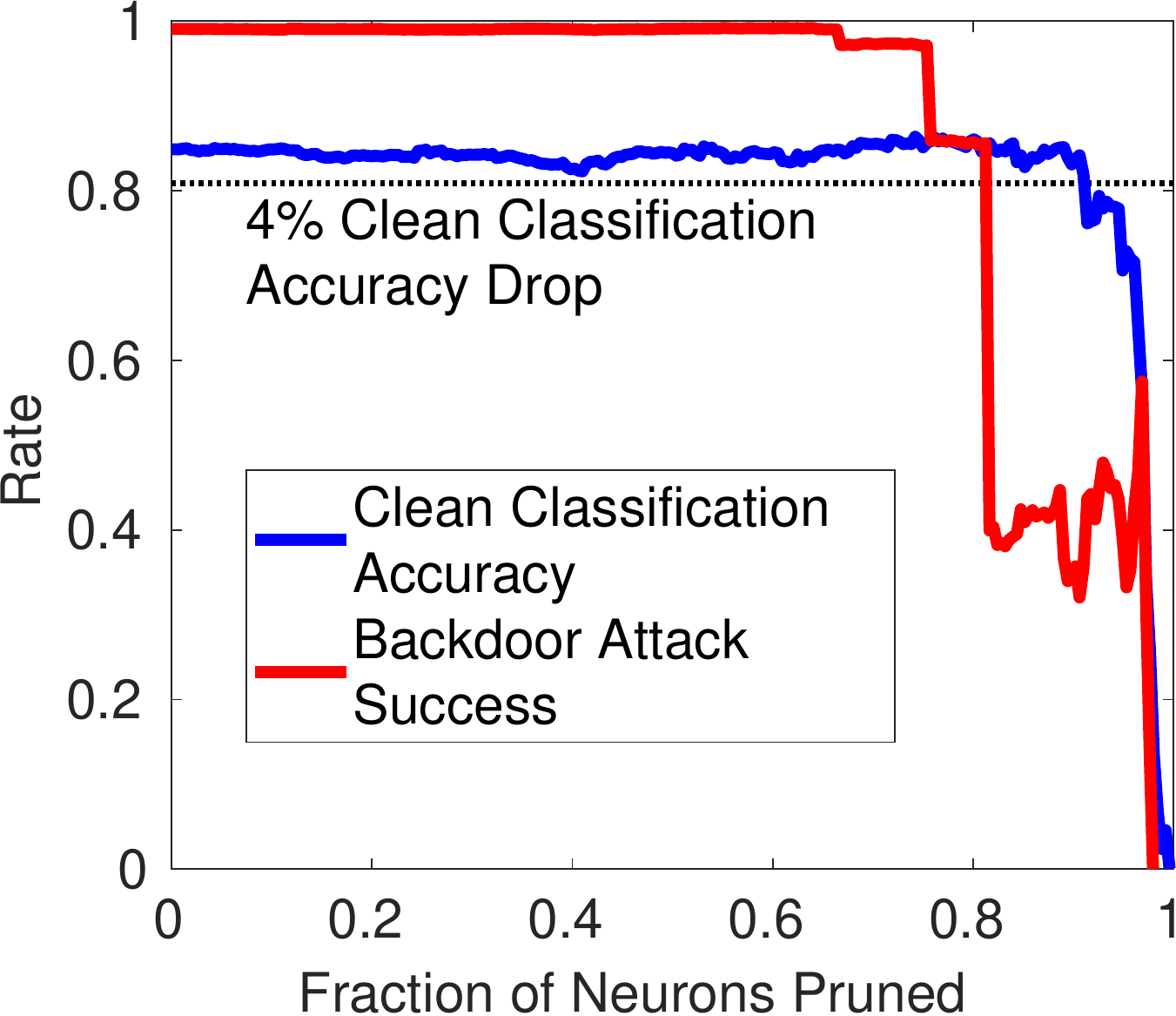}}
    \subfigure[Pruning Aware Attack (Traffic)]{\label{fig:acti_bd_ts}\includegraphics[width=0.45\textwidth,height=1.85in]{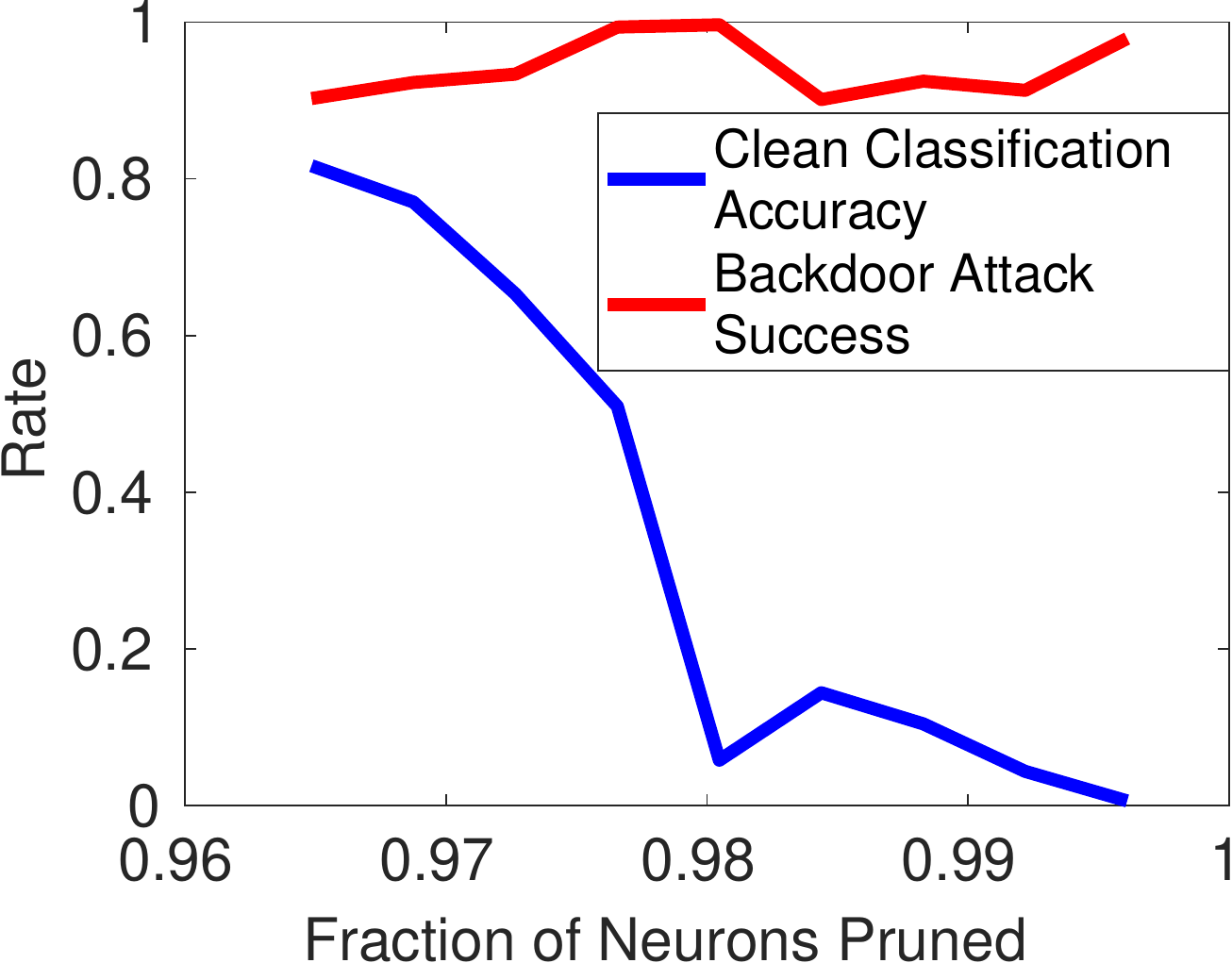}}
    \caption{(a),(c),(e): Classification accuracy on clean inputs and backdoor attack success rate versus fraction of neurons pruned for baseline backdoor attacks on face (a), speech (c) and traffic sign recognition (e). (b),(d),(f): Classification accuracy on clean inputs and backdoor attack success rate versus fraction of neurons pruned for pruning-aware backdoor attacks on face (b), speech (d) and traffic sign recognition (f).  }
    \label{fig:pruning}
\end{figure}

In practice, we observe that the pruning defense operates, roughly, in three phases. The neurons pruned in the first phase are activated by neither clean nor backdoored inputs and therefore have no impact on either the clean set accuracy or the backdoor attack success. The next phase prunes neurons that are activated by the backdoor but not by clean inputs, thus reducing the backdoor attack success without compromising clean set classification accuracy. 
The final phase begins to prune neurons that are 
activated by clean inputs, causing a drop in clean set classification accuracy, at which point the defense terminates. 
These three phases can be seen in Figure~\ref{fig:bs_pr_face}, \ref{fig:bs_pr_speech}, and \ref{fig:bs_pr_ts}.

\paragraph{Empirical Evaluation of Pruning Defense:} 

We evaluated the pruning defense on the face, speech and traffic sign recognition attacks described in Section~\ref{sec:bg:subsec:attacks}. Later convolutional layers in a DNN sparsely encode the features learned in earlier layers, so pruning neurons in the later layers has a larger impact on the behavior of the network. Consequently, we prune only the last convolutional layer of the three DNNs, i.e., conv3 for the DeepID network used in face recognition, conv5 for AlexNet and F-RCNN used in speech and traffic sign recognition, respectively.\footnote{Consistent with prior work, we say ``pruning a neuron'' to mean reducing the number of output channels in a layer by one.}

Figure~\ref{fig:pruning} plots the classification accuracy on clean inputs and 
the success rate of the attack as a function of the number of neurons pruned from the last convolutional layer. Several observations can be made from the figures:
\begin{itemize}
    \item In all three cases, we observe a sharp decline in backdoor attack success rate once sufficiently many neurons are pruned. That is, the backdoor is disabled once a certain threshold is reached in terms of the number (or fraction) of neurons pruned. 
    \item While threshold at which the backdoor attack's success rate drops varies from $0.68\times$ to $0.82\times$ the total number of neurons, the classification accuracy of the pruned networks on clean inputs remains close to that of the original network at or beyond the threshold. Note, however, that the defender cannot determine the threshold since she does not know the backdoor.
    \item Terminating the defense once the classification accuracy on clean inputs drops by more than $4\%$ yields pruned DNNs that are immune to backdoor attacks. Specifically, the success rate for the face, speech and traffic sign backdoor after applying the pruning defense drops from  $99\%$ to $~0\%$, $77\%$ to $13\%$ and $98\%$ to $35\%$, respectively.
\end{itemize}

\paragraph{Discussion:} The pruning defense has several appealing properties from the defender's standpoint. 
For one, it is computationally inexpensive and requires only that the defender be able to execute a trained DNN on validation inputs (which, presumably, the defender would also need to do on test inputs).  
Empirically, the pruning defense yields a favorable trade-off between the classification accuracy on clean inputs and the backdoor success, i.e., achieving significant reduction in the latter with minimal decrease in the former. 

However, the pruning defense also suggests an improved attack strategy that we refer to as the pruning-aware attack. This new strategy is discussed next. 

\subsection{Pruning-Aware Attack}
\label{sec:meth:ss:pruningaware}

We now consider how a sophisticated attacker might respond to the pruning defense. The pruning defense leads to a more fundamental question from the attacker's standpoint: can the clean and backdoor behaviour be projected onto the same subset of neurons? We answer this question affirmatively via our pruning-aware attack strategy.

\begin{figure}
    \centering
    \includegraphics[width=\textwidth]{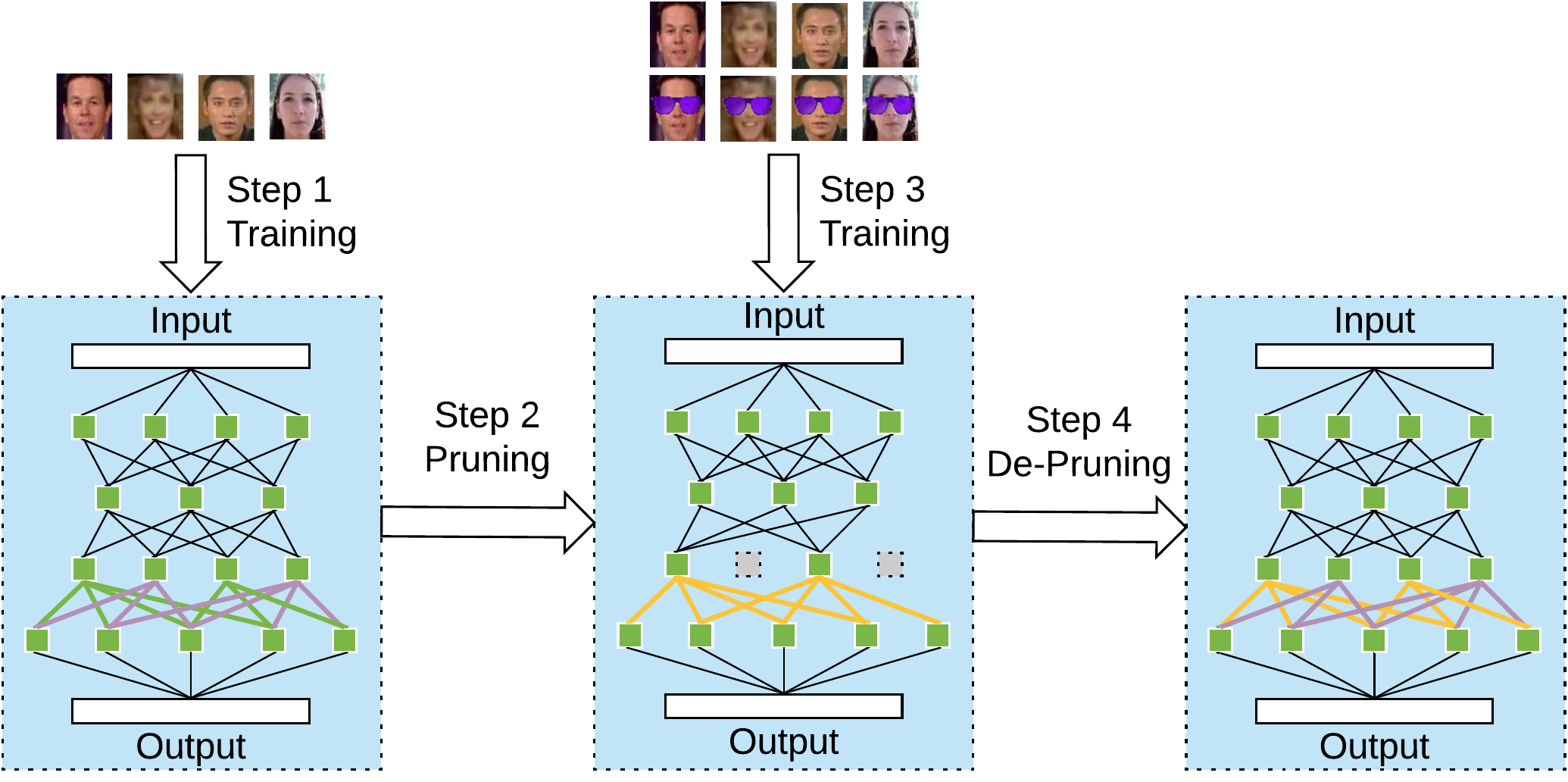}
    \caption{Operation of the pruning-aware attack.}
    \label{fig:pruning_aware_attack}
\end{figure}

The pruning aware attack strategy operates in four steps, as shown in Figure~\ref{fig:pruning_aware_attack}.
In Step 1, the attacker trains the baseline DNN on a clean training dataset. In Step 2, the attacker prunes the DNN by eliminating dormant neurons. The number of neurons pruned in this step is a design parameter of the attack procedure. 
In Step 3, the attacker re-trains the pruned DNN, but this time with the \emph{poisoned} training dataset. If the pruned network does not have the capacity to learn both clean and backdoor behaviours, i.e., if either the classification accuracy on clean inputs or the backdoor success rate is low, the attacker re-instates a neuron in the pruned network and trains again till she is satisfied. 

At the end of Step 3, the attacker obtains a pruned DNN the implements both the desired behaviour on clean inputs \emph{and} the misbehaviour on backdoored inputs.
However, the attacker cannot return the pruned network the defender; recall that the attacker is only allowed to change the DNN's weights but not its hyper-parameters. 
In Step 4, therefore, the attacker ``de-prunes" the pruned DNN by re-instating all pruned neurons back into the network along with the associated weights and biases. However, the attacker must ensure that the re-instated neurons remain dormant on clean inputs; this is achieved by decreasing the biases of the reinstated/de-pruned neurons ($b_i$ in Equation~\ref{eq:dnn-layer}). Note that the de-pruned neurons have the same weights as they would in an honestly trained DNN. Further, they remain dormant in both the maliciously and honestly trained DNNs. Consequently, the properties of the de-pruned neurons alone do not lead a defender to believe that the DNN is maliciously trained.      

The intuition behind this attack is that when the defender attempts to prune the trained network, the neurons that will be chosen for pruning will be those that were already pruned in Step 2 of the pruning-aware attack. Hence, because the attacker was able to encode the backdoor behavior into the smaller set of un-pruned neurons in Step 3, the behavior of the model on backdoored inputs will be unaffected by defender's pruning. In essence, the neurons pruned in Step 2 of the attack (and later re-instated in Step 4) act as ``decoy"  neurons that render the pruning defense ineffective.

\paragraph{Empirical Evaluation of Pruning-Aware Attack:}

\begin{figure}
    \centering
    \subfigure[Clean Activations (pruning aware attack)]{\label{fig:acti_cl_face_dep}\includegraphics[width=0.45\textwidth]{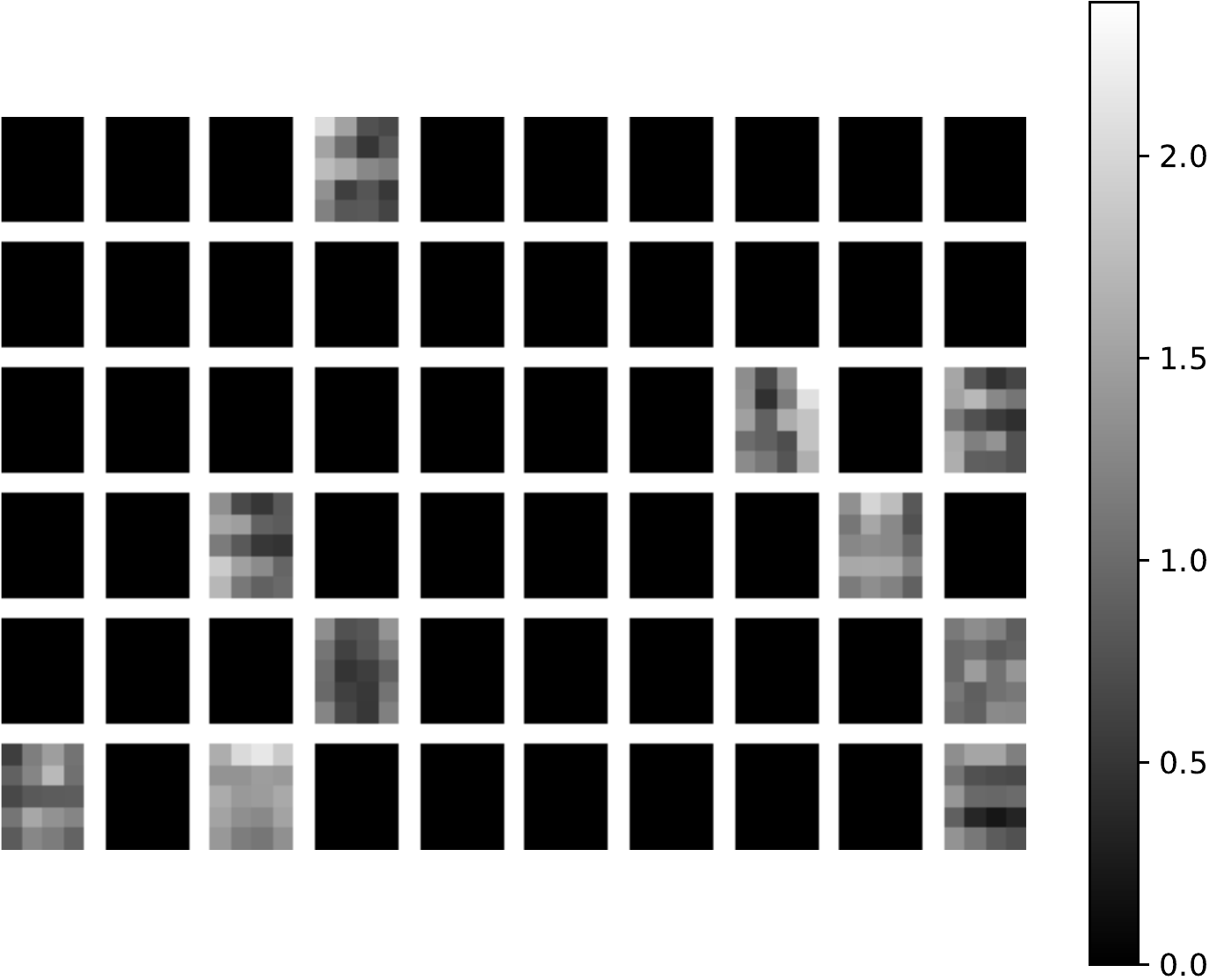}}
    \subfigure[Backdoor Activations (pruning aware attack)]{\label{fig:acti_bd_face_dep}\includegraphics[width=0.45\textwidth]{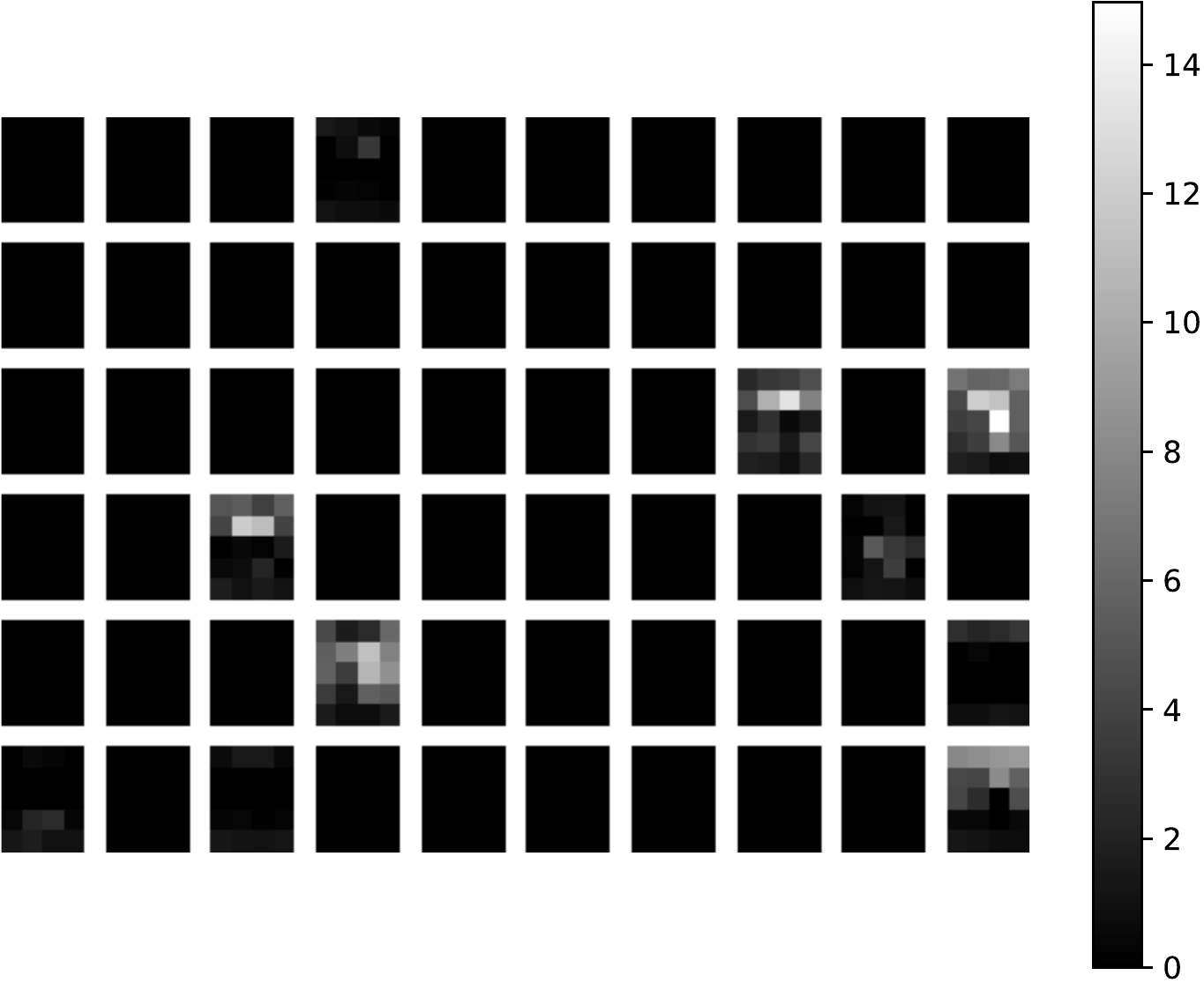}}
    \caption{Average activations of neurons in the final convolutional layer of the backdoored face recognition DNN for clean and backdoor inputs, respectively. The DNN is backdoored using the pruning-aware attack.}
    \label{fig:face-acts-pa}
\end{figure}

Figure~\ref{fig:face-acts-pa} shows the average activations of the last convolutional layer for 
the backdoored face recognition DNN generated by the pruning-aware attack. Note that compared to the activations of the baseline attack (Fig~\ref{fig:face-acts}) 
(i) a larger fraction of neurons remain dormant (about $84\%$) for both clean and backdoored inputs; and (ii) the activations of clean and backdoored inputs are confined to the \emph{same} subset of neurons. Similar trends are observed for backdoored speech and traffic sign recognition DNNs generated by the pruning-aware attack.  
Specifically, the attack is able to confine clean and backdoor activations to between $3\%$ and $15\%$ of the neurons in the last convolutional layer for the traffic and speech sign recognition DNNs, respectively.   

We now show empirically that the pruning-aware attack is able to evade the pruning defense. 
Figure~\ref{fig:pruning}(b),(d),(f) plots the classification accuracy on clean inputs and 
backdoor attack success rate versus the fraction of neurons pruned by the defender for the face, speech and traffic sign recognition networks. 
Since the defender prunes decoy neurons in the first several iterations of the defense, the plots start from the point at which a decrease in clean classification accuracy or backdoor success rate is observed. 

Several observations can be made from the figures:
\begin{itemize}
    \item The backdoored DNNs generated by the baseline and pruning-aware attack have the same classification accuracy on clean inputs assuming a na\"ive defender who does not perform any pruning. This is true for the face, speech and traffic sign recognition attacks.
    \item Similarly, the success rate of the baseline and pruning-aware attack on face and speech recognition are the same, assuming a na\"ive defender who does not perform any pruning. The success rate of the pruning-aware attack reduces slightly to $90\%$ from $99\%$ for the baseline attack for traffic sign recognition, again assuming a na\"ive defender. 
    \item The pruning defense on the backdoored face recognition DNN (see Fig~\ref{fig:pruning}(b)) causes, at a first, in a drop in the classification accuracy on clean inputs but not in the backdoor attack success rate. Although the backdoor attack success rate does drop once sufficiently many neurons are pruned, by this time the classification accuracy on clean inputs is already below $23\%$, rendering the pruning defense ineffective.
    \item The pruning defense on the backdoored speech recognition DNN (see Fig~\ref{fig:pruning}(d)) causes both the classification accuracy on clean inputs \emph{and} the backdoor attacks success rate to gradually fall as neurons are pruned. Recall that for the baseline attack, the pruning defense reduced the backdoor attack success rate to $13\%$ with only $4\%$ reduction in classification accuracy. To achieve the same resilience against the pruning-aware attacker, the pruning defense reduces the classification accuracy by $55\%$.
    \item The pruning defense is also ineffective on backdoored traffic sign recognition (see Fig~\ref{fig:pruning}(f)). Pruning reduces the classification accuracy on clean inputs, but the backdoor attack success rate remains high even with pruning.
\end{itemize}

\paragraph{Discussion:} 
The pruning-aware attack shows that it is not necessary for clean and backdoor inputs to activate different parts of a DNN as observed in prior work~\cite{badnets}. 
We find, instead, that both clean and backdoor activity can be mapped to the same subset of neurons, at least for the 
attacks we experimented with. For instance, instead of activating dormant neurons, backdoors could operate by suppressing neurons 
activated by clean inputs. In addition, the commonly used ReLU activation function, used in all of the DNNs we evaluated in this paper, 
enables backdoors to be encoded by how strongly a neuron is activated as opposed to which neurons are activated since its output ranges from $[0, \infty)$.

\subsection{Fine-Pruning Defense}

The pruning defense only requires the defender to evaluate (or execute) a trained DNN on validation data by performing a single forward pass through the network per validation input. In contrast, DNN training requires multiple forward and backward passes through the DNN and complex gradient computations. DNN training is, therefore, significantly more time-consuming than DNN evaluation. 
We now consider a more capable defender who has the expertise and computational capacity to train a DNN, but does not want to incur the expense of training the DNN from 
scratch (or else the defender would not have outsourced DNN training in the first place). 

Instead of training the DNN from scratch, a capable defender can instead \emph{fine-tune} the DNN trained by the attacker using clean inputs. Fine-tuning is a strategy originally proposed in the context of transfer learning~\cite{yosinski2014transferable}, wherein a user wants to adapt a DNN trained for a certain task to perform another related task. 
Fine-tuning uses the pre-trained DNN weights to initialize training (instead of random initialization) and a smaller learning rate since the final weights are expected to be
relatively close to the pre-trained weights. Fine-tuning is significantly faster than training a network from scratch; for instance, our fine-tuning experiments on AlexNet terminate within an hour while training AlexNet from scratch can take more than six days~\cite{iandola2016firecaffe}. Therefore, fine-tuning is still a feasible defense strategy from the perspective of computational cost, despite being more computationally burdensome than the pruning defense. 

Unfortunately, as shown in Table~\ref{tab:allresults}, the fine-tuning defense does not always work on backdoored DNNs trained using the baseline attack. 
The reason for this can be understood as follows: the accuracy of the backdoored DNN on clean inputs does not depend on the weights of backdoor neurons since these are 
dormant on clean inputs in any case. Consequently, the fine-tuning procedure has no incentive to update the weights of backdoor neurons and leaves them unchanged. 
Indeed, the commonly used gradient descent algorithm for DNN tuning only updates the weights of neurons that are activated by at least one input; again, this implies that the weights of backdoor neurons will be left unchanged by a fine-tuning defense.

\paragraph{Fine-pruning:} The fine-pruning defense seeks to combine the benefits of the pruning and fine-tuning defenses. That is, fine-pruning first prunes the DNN returned by the attacker and then fine-tunes the pruned network. 
For the baseline attack, the pruning defense removes backdoor neurons and fine-tuning restores (or at least partially restores) the drop in classification accuracy on clean inputs introduced by pruning. 
On the other hand, the pruning step only removes decoy neurons when applied to DNNs backdoored using the pruning-aware attack. 
However, subsequent fine-tuning eliminates backdoors. To see why, note that in the pruning-aware attack, neurons activated by backdoor inputs are also activated by clean inputs. Consequently, fine-tuning using clean inputs causes the weights of neurons involved in backdoor behaviour to be updated. 

\paragraph{Empirical Evaluation of Fine-Pruning Defense:}

\begin{table}[]
\centering
\caption{Classification accuracy on clean inputs (cl) and backdoor attack success rate (bd) using fine-tuning and fine-pruning defenses against the baseline and 
pruning-aware attacks.}
\label{tab:allresults}
\begin{tabular}{|l|c|c|c|c|c|c|}
\hline
\multirow{3}{*}{\begin{tabular}[c]{@{}l@{}}Neural \\ Network\end{tabular}} & \multicolumn{3}{c|}{Baseline Attack} & \multicolumn{3}{c|}{Pruning Aware Attack} \\ \cline{2-7} 
 & \multicolumn{3}{c|}{Defender Strategy} & \multicolumn{3}{c|}{Defender Strategy} \\ \cline{2-7} 
 & None & Fine-Tuning & Fine-Pruning & None & Fine-Tuning & Fine-Pruning \\ \hline
\begin{tabular}[c]{@{}l@{}}Face\\ Recognition\end{tabular} & \begin{tabular}[c]{@{}c@{}}cl: 0.978\\ bd: 1.000\end{tabular} & \begin{tabular}[c]{@{}c@{}}cl: 0.978\\ bd: 0.000\end{tabular} & \begin{tabular}[c]{@{}c@{}}cl: 0.978\\ bd: 0.000\end{tabular} & \begin{tabular}[c]{@{}c@{}}cl: 0.974\\ bd: 0.998\end{tabular} & \begin{tabular}[c]{@{}c@{}}cl: 0.978\\ bd: 0.000\end{tabular} & \begin{tabular}[c]{@{}c@{}}cl: 0.977\\ bd: 0.000\end{tabular} \\ \hline
\begin{tabular}[c]{@{}l@{}}Speech\\ Recognition\end{tabular} & \begin{tabular}[c]{@{}c@{}}cl: 0.990\\ bd: 0.770\end{tabular} & \begin{tabular}[c]{@{}c@{}}cl: 0.990\\ bd: 0.435\end{tabular} & \begin{tabular}[c]{@{}c@{}}cl: 0.988\\ bd: 0.020\end{tabular} & \begin{tabular}[c]{@{}c@{}}cl: 0.988\\ bd: 0.780\end{tabular} & \begin{tabular}[c]{@{}c@{}}cl: 0.988\\ bd: 0.520\end{tabular} & \begin{tabular}[c]{@{}c@{}}cl: 0.986\\ bd: 0.000\end{tabular} \\ \hline
\begin{tabular}[c]{@{}l@{}}Traffic Sign\\ Detection\end{tabular} & \begin{tabular}[c]{@{}c@{}}cl: 0.849\\ bd: 0.991\end{tabular} & \begin{tabular}[c]{@{}c@{}}cl: 0.857\\ bd: 0.921\end{tabular} & \begin{tabular}[c]{@{}c@{}}cl: 0.873\\ bd: 0.288\end{tabular} & \begin{tabular}[c]{@{}c@{}}cl: 0.820\\ bd: 0.899\end{tabular} & \begin{tabular}[c]{@{}c@{}}cl: 0.872\\ bd: 0.419\end{tabular} & \begin{tabular}[c]{@{}c@{}}cl: 0.874\\ bd: 0.366\end{tabular} \\ \hline
\end{tabular}
\end{table}

We evaluate the fine-pruning defense on all three backdoor attacks under both the baseline attacker as well as the more sophisticated pruning-aware attacker described in Section~\ref{sec:meth:ss:pruningaware}. The results of these experiments are shown under the ``fine-pruning'' columns of Table~\ref{tab:allresults}. We highlight three main points about these results:
\begin{itemize}
    \item In the worst case, fine-pruning reduces the accuracy of the network on clean data by just 0.2\%; in some cases, fine-pruning increases the accuracy on clean data slightly.
    \item For targeted attacks, fine-pruning is highly effective and completely nullifies the backdoor's success in most cases, for both the baseline and pruning-aware attacker. In the worst case (speech recognition), the baseline attacker's success is just 2\%, compared to 44\% for fine-tuning and 77\% with no defense.
    \item For the untargeted attacks on traffic sign recognition, fine-pruning reduces the attacker's success from 99\% to 29\% in the baseline attack and from 90\% to 37\% in the pruning-aware attack. Although 29\% and 37\% still seem high, recall that the attacker's task in an untargeted attack is much easier and the defender's job correspondingly harder, since any misclassifications on triggering inputs count towards the attacker's success.
\end{itemize}

\paragraph{Discussion:}

Given that both fine-pruning and fine-tuning work equally well against a pruning-aware attacker, one may be tempted to ask why fine-pruning is needed. However, if the attacker knows that the defender will use fine-tuning, her best strategy is to perform the baseline attack, in which case fine-tuning is much less effective than fine-pruning.

\begin{table}[]
\centering
\caption{Defender's utility matrix for the speech recognition attack. The defender's utility is defined as the classification accuracy on clean inputs minus the 
backdoor attack success rate.}
\label{tab:speech_util}
\begin{tabular}{|c|l|c|c|}
\hline
\multicolumn{2}{|c|}{\multirow{2}{*}{Utility}} & \multicolumn{2}{c|}{Attacker Strategy} \\ \cline{3-4} 
\multicolumn{2}{|c|}{} & Baseline Attack & Pruning Aware Attack \\ \hline
\multirow{2}{*}{\begin{tabular}[c]{@{}c@{}}Defender\\ Strategy\end{tabular}} & Fine-Tuning & 0.555 & 0.468 \\ \cline{2-4} 
 & Fine-Pruning & 0.968 & 0.986 \\ \hline
\end{tabular}
\end{table}

One way to see this is to consider the \emph{utility matrix} for a baseline and pruning-aware attacker against a defender using fine-tuning or fine-pruning. The utility matrix for the speech recognition attack is shown in Table~\ref{tab:speech_util}. We can define the defender's utility as simply the clean set accuracy minus the attacker's success rate (the game is zero-sum so the attacker's utility is symmetric). From this we can see that defender's best strategy is \emph{always} to use fine-pruning.
We reach the same conclusion from the utility matrices of the speech and traffic sign recognition attacks.

Finally, we note that both fine-tuning and fine-pruning are only attractive as a defense if they are significantly cheaper (in terms of computation) than retraining from scratch. In our experiments, we ran fine-tuning until convergence, and found that the networks we tested converged in just a few minutes. Although these experiments were performed on a cluster with high-end GPUs available (NVIDIA P40, P100, K80, and GTX 1080), even if a less powerful GPU is used (say, one that is 10X slower) we can see that fine-pruning is still significantly more efficient than training from scratch, which can take several \emph{days} in the case of large models such as AlexNet~\cite{iandola2016firecaffe}.

\section{Discussion}

Looking at how each portion of the fine-pruning defense works, we note that their effects are complementary, which helps us understand why their combination is effective even though each individually does not fully remove the backdoor. Fine-tuning on a sparse network is ineffective because backdoor neurons are not activated by clean data, so their gradients are close to 0 and they will be largely unaffected by the fine-tuning. However, these are precisely the neurons that will be selected for pruning, since their activations on clean data are low. It is only once we prune \emph{and} fine-tune, forcing the attacker to concentrate her backdoor into a relatively small number of neurons, that fine-tuning can act on neurons that encode the backdoor trigger.

The fact that backdoors can be removed automatically is surprising from the perspective of prior research into backdoors in traditional software and hardware. Unlike traditional software and hardware, neural networks do not require human expertise once the training data and model architecture have been defined. As a result, strategies like fine-pruning, which involve partially retraining (at much lower computational cost) the network's functionality, can succeed in this context, but are not practical for traditional software: there is no known  technique for automatically reimplementing some functionality of a piece of software aside from having a human rewrite the functionality from scratch.

We cannot guarantee that our defense is the last word in DNN backdoor attacks and defenses. We can think of the fine-tuning as a continuation of the normal training procedure from some set of initialization parameters $\Theta_{i}$. In an adversarial context, $\Theta_{i}$ is determined by the attacker. Hence, if an attacker hopes to preserve their attack against our fine-pruning, they must provide a $\Theta_{i}$ with a nearby local minimum (in terms of the loss surface with respect to the clean dataset) that still contains their backdoor. We do not currently have a strong guarantee that such a $\Theta{i}$ \emph{cannot} be found; however, we note that a stronger (though more computationally expensive) version of fine-pruning could add some noise to the parameters before fine-tuning. In the limit, there must exist some amount of noise that would cause the network to ``forget'' the backdoor, since adding sufficiently large amounts of noise would be equivalent to retraining the network from scratch with random initialization. We believe the question of how much noise is needed to be an interesting area for future research.

\subsection{Threats to Validity}

The backdoor attacks studied in this paper all share a similar underlying model architecture: convolutional neural networks with ReLU activations. These networks are widely used in practice for many different tasks, but they are not the only architectures available. For example, recurrent neural networks (RNNs) and long short term memory networks (LSTMs) are commonly used in sequential processing tasks such as natural language processing. To the best of our knowledge, backdoor attacks have not yet been explored thoroughly in these architectures; as a result, we cannot be sure that our defense is applicable to all deep networks.

\section{Related Work}

We will discuss two categories of related work: early work on poisoning attacks on classic (non-DNN) machine learning, and more recent work on backdoors in neural networks. We will not attempt to recap, here, the extensive literature on adversarial inputs and defenses so far. Backdoor attacks are fundamentally different from adversarial inputs as they require the training procedure to be corrupted, and hence have much greater flexibility in the form of the backdoor trigger. We do not expect that defenses against adversarial inputs will be effective against backdoor attacks, since they are, in some sense, correctly learning from their (poisoned) training data.
 
Barreno et al.~\cite{Barreno:2006} presented a useful taxonomy for classifying different types of attacks on machine learning along three axes: whether the goal is to compromise the \emph{integrity} or \emph{availability} of the system, whether the attack is \emph{exploratory} (gaining information about a trained model) or \emph{causative} (changing the output of the model by interfering with its training data), and whether the attack is \emph{targeted} or \emph{indiscriminate}.

Many of the early attacks on machine learning were \emph{exploratory} attacks on network and host-based intrusion detection systems~\cite{Wagner:2002,Tan:2002,Fogla:2006,Fogla:2006a} or spam filters~\cite{Wittel:2004,Lowd:2005,lowd2005good,Karlberger:2007}. Causative attacks, primarily using \emph{training data poisoning}, soon followed, again targeting spam filtering~\cite{Nelson:2008} and network intrusion detection~\cite{Chung:2006,Chung:2007,Newsome:2006}. Many of the  these attacks focused on systems which had some \emph{online learning} component in order to introduce poisoned data into the system. Suciu et al.~\cite{suciu2018does} classify poisoning and evasion attacks into a single framework for modeling attackers of machine learning systems, and present StingRay, a targeted poisoning attack that is effective against several different machine learning models, including convolutional neural networks. Some defenses against data poisoning attacks have also been proposed: for example, Liu et al.~\cite{liu2017regression} discuss a technique for performing robust linear regression in the presence of noisy data and adversarially poisoned training samples by recovering a low-rank subspace of the feature matrix.

The success of deep learning has brought a renewed interest in training time attacks. Because training is more expensive, outsourcing is common and so threat models in which the attacker can control the parameters of the training procedure are more practical. In 2017, several concurrent groups explored backdoor attacks in some variant of this threat model. In addition to the three attacks described in detail in Section~\ref{sec:bg:subsec:attacks}~\cite{badnets,berkeley,Trojannn}, Mu\~noz-Gonz\'alez et al.~\cite{biggio:2017} described a gradient-based method for producing poison data, and Liu et al.~\cite{Liu:2017} examine \emph{neural trojans} on a toy MNIST example and evaluate several mitigation techniques. In the context of the taxonomy given by Barreno et al.~\cite{Barreno:2006}, these backdoor attacks can be classified as causative integrity attacks.

Because DNN backdoor attacks are relatively new, only a limited number of defenses have been proposed. Chen et al.~\cite{berkeley} examine several possible countermeasures, including some limited retraining with a held-out validation set, but conclude that their proposed defenses are ineffective. Similarly, in their NDSS 2017 paper, Liu et al.~\cite{Trojannn} note that targeted backdoor attacks will disproportionately reduce the accuracy of the model on the targeted class, and suggest that this could be used as a detection technique. Finally, Liu et al.'s~\cite{Liu:2017} mitigations have only been tested on the MNIST task, which is generally considered unrepresentative of real-world computer vision tasks~\cite{xiao2017}. Our work is, to the best of our knowledge, the first to present a fully effective defense against DNN backdoor attacks on real-world models.

\section{Conclusion}

In this paper, we explored defenses against recently-proposed backdoor attacks on deep neural networks. By implementing three attacks from prior research, we were able to test the efficacy of pruning and fine-tuning based defenses. We found that neither provides strong protection against backdoor attacks, particularly in the presence of an adversary who is aware of the defense being used. Our solution, \emph{fine-pruning}, combines the strengths of both defenses and effectively nullifies backdoor attacks. Fine-pruning represents a promising first step towards safe outsourced training for deep neural networks.

\bibliographystyle{abbrv}
\bibliography{main,fap}

\end{document}